\newcommand\approxlt{\mbox{$^{<}\hspace{-0.24cm}_{\sim}$}}
\newcommand\approxgt{\mbox{$^{>}\hspace{-0.24cm}_{\sim}$}}

\documentclass{aa}
\usepackage{psfig,latexsym,longtable,lscape}

\def\ros{{\sl ROSAT }}
\def\sax{{\sl Beppo-SAX }}

\def\asec{\ifmmode ^{\prime\prime}\else$^{\prime\prime}$\fi}
\def\amin{\ifmmode ^{\prime}\else$^{\prime}$\fi}

\begin{document}

   \thesaurus{6(11.09.1;08.02.1;13.25.5;08.05.3;08.23.1)} 
 \title{The M31 population of supersoft sources}

   \author{P. Kahabka\inst{}}

   \offprints{ptk@astro.uva.nl}
 
   \institute{Astronomical Institute and Center for High Energy
   Astrophysics, University of Amsterdam, Kruislaan 403, 1098~SJ 
   Amsterdam, The Netherlands}

   \date{Received 9 February 1998 / Accepted 11 January 1999}

  \titlerunning{The M31 population of Supersoft Sources}
  \authorrunning{P. Kahabka}
  \maketitle
 
   \begin{abstract}
   The 1991 \ros {\sl PSPC} M31 X-ray point source catalog has been screened
   in order to set up a sample of candidate supersoft sources in this galaxy,
   additional to the 16 supersoft sources already known in M31 (Supper et al.
   1997). Selection criteria used were based on hardness ratios (``X-ray 
   colors''), as developed in an earlier paper (Kahabka 1998). An additional 
   criterion to be fulfilled was that the observed count rate is in agreement 
   with the expected steady-state luminosity for a source with these hardness 
   ratios. This condition constrained mainly the hydrogen absorbing column 
   towards the source. 26 candidates not correlating with foreground stars 
   and M31 supernova remnants have been found to fulfil one of the selection 
   criteria. They can be considered to be candidate supersoft sources in M31. 
   This comprises 6\% of all point sources in this galaxy. For these 
   candidates absorbing hydrogen column densities, effective temperatures and 
   white dwarf masses (assuming the sources are on the stability line of 
   surface nuclear burning) are derived. An observed white dwarf mass 
   distribution is derived which indicates that the masses are constrained 
   to $M\approxgt$ 0.90 $\ M_{\odot}$. 

   The entire population of supersoft sources in M31 is estimated taking a 
   theoretical white dwarf mass distribution into account and under the 
   assumption that the observationally derived sample is restricted to white 
   dwarf masses above $0.90\ M_{\odot}$. Taking into account that the gas and
   the source population have different scale heights a total number of at 
   least 200-500 and at most 6,000-15,000 sources is deduced (depending on the
   used galaxy $N_{\rm H}$ model), making use of the population synthesis 
   calculation of Yungelson (1996).

   The spatial distribution favors a disk (or spiral-arm) dominated young 
   stellar population with a ratio of 1/(4-7) of bulge/disk systems, very 
   similar to what has been found for novae in the Milky Way but lower than 
   favored for novae in M31 ($\sim$1/2). Supersoft sources and 
   Cepheids both show association with the M31 spiral arms and may belong to 
   a younger stellar population. A mean space density of $\sim (0.1-5)\times
   10^{-8}\ {\rm pc^{-3}}$ is inferred for the supersoft sources. Assuming 
   that all supersoft sources with masses in excess of $0.5\ M_{\odot}$ are 
   progenitors of supernovae of type Ia, a SN~Ia rate of $(0.8-7)\times 
   10^{-3}\ {\rm yr^{-1}}$ is derived for M31 based on these progenitors. 
   Supersoft sources might be able to account for 20-100\% of the total SN~Ia 
   rate in a galaxy like M31.

%______________________________________ Do not leave a blank line here!

%  14.Sep.'90: Demo-Vs.
%_____________________________________ Do not leave a blank line here!
 
   \end{abstract}
%
%  14.Sep.'90: Demo-Vs.
%________________________________________________________________
 
\section{Introduction}
Supersoft sources constitute an interesting new class of X-ray binary sources 
(cf. van den Heuvel et al. 1992). This is due to the fact that their spectra 
are extremely soft (effective temperatures of a few $10^5\ {\rm K}$) and their
luminosities are substantial ($10^{36}-10^{38}\ {\rm erg\ s^{-1}}$). They thus
can not only be studied in the Milky Way and the near-by Magellanic Clouds 
(LMC and SMC) but also in more distant galaxies (the Andromeda galaxy M31 and 
the spiral galaxy NGC~55). For a review see Hasinger (1994), Kahabka \& 
Tr\"umper (1996) and Kahabka \& van den Heuvel (1997), see also Greiner 
(1996). They are considered to be at least one class of progenitors of type 
Ia supernovae (cf. Branch et al.1995, Livio 1996, Li \& van den Heuvel 1997,
Yungelson \& Livio 1998, and Branch 1998). 

In this article the observed sample of supersoft sources in the Andromeda 
galaxy (M31) is studied. A distance of 700~kpc is adopted (slightly different 
distances of 690 and 725~kpc are used in other literature). The inclination 
of the galaxy is 77.5$^o$. The inner ``apparent bulge'' has a radius of 3~kpc 
and the bulge is truncated at a radius of 6.4~kpc. The disk extends to a 
radius of $\sim$20~kpc (Hatano et al. 1997). First a number of candidate 
supersoft sources in M31 is derived from the 1991 X-ray point source catalog 
(retrieved from CDS via anonymous ftp 130.79.128.5) making use of the hardness
ratios (``X-ray colors'') HR1, HR2 and the count rate information. 

For the 26 candidates a hydrogen column ($N_{\rm H}$) distribution and a 
white dwarf mass distribution is derived assuming non-LTE white dwarf 
atmosphere spectra. The observationally derived white dwarf mass distribution 
is compared with the mass distribution predicted from population synthesis 
calculations and the number of the population is corrected accordingly. The 
observationally derived $N_{\rm H}$ distribution is compared with a galaxy 
scale height $N_{\rm H}$ distribution and the population is corrected 
accordingly. This allows to constrain the whole active population distributed 
over the whole galaxy. Assuming objects with masses $>0.8\ M_{\odot}$ (and 
$>0.5\ M_{\odot}$ respectively) contribute as progenitors of type Ia SNe and 
explode after an evolutionary time scale of $\sim10^6\ yr$ a SN Ia rate is 
derived for M31 from this population. This rate is compared with the total M31
SN Ia rate.

%
%  14.Sep.'90: Demo-Vs.
%__________________________________________________________________

\section{The sample of supersoft sources in M31}
\subsection{The SW-sample}
15 firm candidate supersoft sources have been found in the 1991 \ros 
{\sl PSPC} observations of M31 by Supper et al. (1997), cf. Greiner, Supper \&
Magnier (1997) by applying to the {\sl ROSAT PSPC} hardness ratio HR1 which 
is defined as 

\begin{equation}{\rm
  HR1=(H-S)/(H+S)}
\end{equation}

with S = counts in channel 11-41 (roughly 0.1-0.4 keV), H = counts in channel 
52-201 (roughly 0.5-2.1 keV). The selection criterion for a supersoft source
is:

\begin{equation}{\rm
  HR1 + \sigma  HR1 \le -0.80}
\end{equation}

A 16-th supersoft source, a recurrent transient has been discovered by White 
et al. (1994). We call this sub-sample of 16 sources the Supper-White
(SW) sample. 

It has been shown by Kahabka (1998) that the individual spectral parameters
giving information on the white dwarf masses of these 16 M31 supersoft sources
can be constrained if the \ros {\sl PSPC} hardness ratios HR1 and HR2 and the 
count rate as given in the catalog of Supper et al. (1997) are taken into 
account. The definition of HR2 is

\begin{equation}{\rm
  HR2=(H2-H1)/(H2+H1)}
\end{equation}

with H1 = counts in channel 52-90 (roughly 0.5-0.9 keV), H2 = counts in 
channel 91-201 (roughly 0.9-2.0 keV). The hardness ratios HR1 and HR2 and the 
count rates have been compared with theoretical values derived using non-LTE 
white dwarf atmosphere spectra. As a result we found that for all these 16
sources the white dwarf masses were quite large $>0.9\ M_{\odot}$.

In the present work non-LTE models of white dwarf atmospheres are used 
(Hartmann, private communication) extending to effective temperatures as low 
as $3\times10^5\ K$. Absorbing hydrogen columns $N_{\rm H}$ and effective 
temperatures $T_{\rm eff}$ have been determined in a $N_{\rm H}$-$T_{\rm eff}$
plane from the overlap of the 90\% 
confidence parameter regions as determined from the HR1, HR2 and count rate 
constraints in that plane. In order to calculate the HR1-effective 
temperature planes somewhat reduced errors (0.85$\times 1\sigma$ errors) 
have been used. This has the effect of bounding the hydrogen column to 
$N_{\rm H} > 8.\times 10^{20}\ {\rm H-atoms\ cm^{-2}}$ which is consistent 
with a minimum $N_{\rm H}\sim 6.\times 10^{20}\ {\rm H-atoms\ cm^{-2}}$ due 
to the galactic foreground 
column. In Table~1 values are given for the SW-sample taking these new 
constraints into account. These values differ only slightly from those derived
in Kahabka (1998). Count rates are given for the broad (0.1-2.4~keV) band. 
It should be noted that although the standard deviations for some sources
are very large (which might suggest that these sources are not detected
significantly) all these sources have been detected significantly in the 
soft (0.1-0.4~keV) band (see Table~5 of Supper et al. 1997). The high
standard deviations in the 0.1-2.4~keV band are due to the fact that in
this band most of the counts are background ones from the 0.4-2.4~keV
range.

White dwarf masses are determined under the assumption that the source is 
on the stability line of surface hydrogen burning (cf. Iben 1982). The 
source may even be on the plateau of the Hertzsprung-Russell diagram with 
radius expansion. Then its mass would be even larger. This method has been 
applied to the \sax observation of CAL87 and CAL83 and reasonable white dwarf 
masses of $\approx 1.2\ M_{\odot}$ and $\approx 0.9-1.0 \ M_{\odot}$ have been 
derived respectively (Parmar et al. 1997, 1998).

\subsection{The complimentary sample (C-sample)}
The SW sample cannot be complete as it has been shown that supersoft sources
are expected to cover a much wider range in HR1 (Kahabka 1998). Actually
all values of HR1 in the range $-1.0\le HR1 \approxlt +0.8$ are possible in
case the hottest (most massive) and more strongly absorbed ($N_{\rm H} \le 
5.\times 10^{21}\ {\rm cm^{-2}}$) sources are included. Sources lying deep 
inside the galaxy 
disk or even located below the galaxy disk are expected to be at least in
part even more strongly absorbed and are not covered by the selection criterion
used for the SW sample. It may well be that part of this population is 
detectable but in order to investigate this point the calculations of model 
atmosphere spectra have to be extended to $N_{\rm H}$ values in excess of the 
present upper bound of $5.\times 10^{21}\ {\rm cm^{-2}}$.

\begin{table*}
  \caption[]{\ros {\sl PSPC} count rates (0.1-2.4~keV), hardness ratios HR1, 
             from non-LTE white dwarf atmosphere models M4 and M5 derived 
             absorbing hydrogen columns $(10^{21}\ {\rm cm^{-2}}$), effective 
             temperatures $T_{\rm eff}\ (10^5\ {\rm K})$, white dwarf masses 
             $M_{\rm WD}\ (M_{\odot})$, index and tentative identification from the 
             Supper (1997) catalog (a = foreground star, e = SNR, * = bulge 
             source).}
  \begin{flushleft}
  \begin{tabular}{clccccccrc}
  \hline
  \hline
  \noalign{\smallskip}
& Source name &rate   & HR1& HR2 &$N_{\rm H}$ & $T_{\rm eff}$  & $M_{\rm WD}$     &Supper   & Remark            \\
&&($10^{-3}\ {\rm s^{-1}})$&&&($10^{21}\ {\rm cm^{-2}})$&($10^5\ {\rm K}$)& ($M_{\odot}$)     &Cat./Id. &                   \\
  \noalign{\smallskip}
  \hline
  \noalign{\smallskip}
\multicolumn{10}{c}{the SW-sample} \\
  \noalign{\smallskip}
  \hline
  \noalign{\smallskip}
SWa &RX~J0037.4+4015&0.31$\pm$0.31&-0.93$\pm$0.31& 0.02$\pm$0.71& 1.4-1.7  & 3.7-4.3& 0.92-0.98&    3   & [3,H] \\
SWb &RX~J0038.5+4014&0.80$\pm$0.28&-0.92$\pm$0.08&-0.49$\pm$0.53& 1.2-1.8  & 3.7-4.5& 0.92-1.00&   12   & [1,H] \\
SWc &RX~J0038.6+4020&1.73$\pm$0.29&-0.93$\pm$0.06& 0.32$\pm$0.66& 1.1-1.4  & 4.0-4.6& 0.96-1.02&   18   & [3,H] \\
SWd &RX~J0039.6+4054&0.44$\pm$0.44&-0.92$\pm$0.02&-0.04$\pm$0.71& 1.3-1.6  & 3.9-4.4& 0.94-0.99&   39   & [2,H] \\
SWe &RX~J0040.4+4009&0.85$\pm$0.32&-0.94$\pm$0.06&-0.90$\pm$0.10& 1.2-1.6  & 3.6-4.4& 0.94-0.99&   78   & [1,H] \\
SWf &RX~J0040.7+4015&1.26$\pm$0.32&-0.94$\pm$0.06&-0.31$\pm$0.64& 1.1-1.5  & 3.7-4.5& 0.92-1.00&   88   & [2,H] \\
SWg &RX~J0041.5+4040&0.32$\pm$0.18&-0.95$\pm$0.05&-0.62$\pm$0.44& 1.4-1.9  & 3.4-4.1& 0.90-0.96&  114   & [1,H] \\
SWh &RX~J0041.8+4059&0.49$\pm$0.24&-0.93$\pm$0.07&-0.63$\pm$0.43& 1.3-1.9  & 3.6-4.4& 0.92-0.99&  128   & [1,H] \\
SWi &RX~J0042.4+4044&1.69$\pm$0.32&-0.93$\pm$0.07&-0.07$\pm$0.70& 1.0-1.4  & 3.8-4.6& 0.94-1.02&  171   & [2,H] \\
SWj &RX~J0043.5+4207&2.15$\pm$0.55&-0.92$\pm$0.08&-0.27$\pm$0.66& 1.0-1.4  & 3.9-4.8& 0.95-1.03&  245   & [1,H] \\
SWk &RX~J0044.0+4118&2.46$\pm$0.42&-0.94$\pm$0.06& 0.11$\pm$0.81& 0.9-1.3  & 3.9-4.7& 0.95-1.02&  268   & [3,H] \\
SWl &RX~J0045.5+4206&3.14$\pm$0.34&-0.89$\pm$0.07&-0.29$\pm$0.65& 1.0-1.3  & 4.4-4.9& 0.99-1.04&  309   & [2,H] \\
SWm &RX~J0046.2+4144&2.15$\pm$0.39&-0.93$\pm$0.07&0.62$\pm0$.40 & 1.2-1.7  & 3.8-4.7& 0.94-1.02&  335   & [2,H] \\
SWn &RX~J0046.2+4138&1.12$\pm$0.40&-0.91$\pm$0.09&-0.27$\pm$0.65& 1.0-1.4  & 3.8-4.7& 0.94-1.02&  341   & [2,H] \\
SWo &RX~J0047.6+4205&1.05$\pm$0.36&-0.92$\pm$0.07& 0.06$\pm$0.70& $\le$1.6 & $\le$4.0& $\le$1.03& 376   & [3]   \\
  \noalign{\smallskip}
  \hline
  \noalign{\smallskip}
SWt &RX~J0045.4+4154&29.6$\pm$1.0&+0.78$\pm$0.03&-0.59$\pm$0.03& 4.0-4.2 & 8.4-8.5& 1.26-1.27  &        & [2,H] \\
  \noalign{\smallskip}
  \hline
  \noalign{\smallskip}
\multicolumn{10}{c}{complimentary sample to the SW-sample (C-sample)} \\
  \noalign{\smallskip}
  \hline
  \noalign{\smallskip}
C17 &RX~J0039.7+4030&2.03$\pm$0.30&-0.85$\pm$0.10&-0.83$\pm$0.53& 1.2-1.6  & 4.4-4.9& 0.99-1.04&   45   & [1,H] \\
C18 &RX~J0047.0+4201&0.90$\pm$0.33&-0.84$\pm$0.22&-0.73$\pm$0.30& 0.8-2.1  & 3.2-4.6& 0.87-1.01&  358/a & [1]   \\
C19 &RX~J0043.9+5148&1.29$\pm$0.36&-0.38$\pm$0.24&-0.87$\pm$0.40& 1.8-3.4  & 4.6-5.3& 1.01-1.06&  259   & [1,H] \\
C20 &RX~J0040.5+4034&1.30$\pm$0.30&-0.37$\pm$0.23&-0.68$\pm$0.29& 1.8-3.4  & 4.6-5.3& 1.01-1.06&   82/a & [1]   \\
C21 &RX~J0041.8+4015&3.18$\pm$0.58&-0.35$\pm$0.14&-0.63$\pm$0.27& 1.7-2.2  & 5.0-5.4& 1.04-1.07&  129   & [2,H] \\
C22 &RX~J0047.0+4157&1.79$\pm$0.43&-0.14$\pm$0.24&-0.77$\pm$0.22& 2.2-4.0  & 5.0-5.9& 1.04-1.11&  356/a & [1,H] \\
C23 &RX~J0039.4+4050&2.93$\pm$0.36&-0.05$\pm$0.12&-0.36$\pm$0.16& 2.6-3.3  & 5.5-5.9& 1.08-1.11&   35   & [2,H] \\
C24 &RX~J0044.4+4200&1.17$\pm$0.31& 0.16$\pm$0.30&-0.58$\pm$0.24& 3.5-5.4  & 5.3-6.5& 1.06-1.15&  280   & [2,H] \\
C25 &RX~J0043.7+4127&1.12$\pm$0.32& 0.21$\pm$0.40&-0.44$\pm$0.33& 3.2-7.0  & 5.2-7.0& 1.11-1.21&  252   & [2,H] \\
C26 &RX~J0042.8+4115&40.14$\pm$1.06&0.28$\pm$0.02&-0.18$\pm$0.03& 1.4-1.5  & 7.4-7.5& 1.21-1.22&  208/* & [2,H] \\
C27 &RX~J0040.0+4100&2.04$\pm$0.32& 0.33$\pm$0.17&-0.27$\pm$0.17& 4.1-5.2  & 6.1-6.7& 1.12-1.17&   58   & [2,H] \\
C28 &RX~J0042.2+4048&0.58$\pm$0.24& 0.36$\pm$0.50&-0.72$\pm$0.45& 4.0-10   & 4.5-7.2& 1.10-1.20&  156   & [1]   \\
C29 &RX~J0046.3+4238&3.10$\pm$0.64& 0.37$\pm$0.24&-0.32$\pm$0.20& 3.4-6.0  & 6.1-7.2& 1.12-1.20&  342   & [2,H] \\
C30 &RX~J0045.4+4219&1.19$\pm$0.33& 0.39$\pm$0.32&-0.52$\pm$0.28& 4.6-5.8  & 5.9-6.9& 1.11-1.18&  307   & [2,H] \\
C31 &RX~J0043.4+4118&6.86$\pm$0.62& 0.48$\pm$0.08&-0.36$\pm$0.09& 3.6-4.2  & 6.9-7.2& 1.18-1.20&  240/e & [3]   \\
C32 &RX~J0043.3+4120&6.74$\pm$0.62& 0.49$\pm$0.08&-0.64$\pm$0.09& 3.8-4.4  & 6.9-7.2& 1.18-1.20&  235   & [2,H] \\
C33 &RX~J0045.2+4136&2.44$\pm$0.45& 0.49$\pm$0.24&-0.41$\pm$0.19& 4.2-6.4  & 6.3-7.4& 1.14-1.21&  297/e & [2]   \\
C34 &RX~J0039.7+4039&0.87$\pm$0.23& 0.50$\pm$0.35&-0.65$\pm$0.23& $>$4.8   & $>$5.9 & $>$1.11  &   44/a & [1]   \\
C35 &RX~J0046.1+4136&0.25$\pm$0.25& 0.51$\pm$0.48&-0.74$\pm$0.27& $>$3.0   & $>$6.0 & $>$1.12  &  330   & [1]   \\
C36 &RX~J0043.6+4126&2.40$\pm$0.38& 0.55$\pm$0.20&-0.77$\pm$0.13& 4.5-6.7  & 6.5-7.5& 1.15-1.21&  249/e & [1]   \\
C37 &RX~J0040.1+4021&0.46$\pm$0.19& 0.62$\pm$0.33&-0.78$\pm$0.28& $>$6.8   & $>$6.0 & $>$1.12  &   62   & [1]   \\
C38 &RX~J0042.6+4043&1.55$\pm$0.31& 0.63$\pm$0.22&-0.97$\pm$0.18& $>$5.6   & $>$6.4 & $>$1.14  &  185   & [1]   \\
C39 &RX~J0042.9+4059&0.90$\pm$0.28& 0.64$\pm$0.31&-0.71$\pm$0.25& $>$5.3   & $>$6.3 & $>$1.14  &  212/e & [1]   \\
C40 &RX~J0047.6+4132&0.32$\pm$0.32& 0.75$\pm$0.26&-0.85$\pm$0.70& $>$4.4   & $>$7.0 & $>$1.19  &  374   & [1]   \\
C41 &RX~J0042.6+4159&1.75$\pm$0.82& 0.85$\pm$0.14&-0.68$\pm$0.25& $>$4.4   & $>$7.4 & $>$1.21  &  183   & [1]   \\
C42 &RX~J0044.2+4026&0.07$\pm$0.07& 0.89$\pm$0.13&-0.84$\pm$0.16& $>$5.0   & $>$8.0 & $>$1.24  &  271   & [1]   \\ 
  \noalign{\smallskip}
  \hline
  \end{tabular}
  \end{flushleft}
  Remarks: quality flag [1] = full overlap of HR1, HR2, CPS 
  contours, [2] = medium overlap of HR1 and CPS contours and overlap of 
  HR2 contours considering $3-\sigma$ uncertainties, [3] = full overlap 
  of HR1 and CPS contours but no overlap of HR2 contour possibly due to 
  source confusion in the hard band for the SW-sample of supersoft sources 
  in M31 (Supper et al. 1997, White et al. 1994, cf. Greiner, Magnier \& 
  Hasinger 1997) and for the C-sample, [H] = histogram flag, entry
  in $N_{\rm H}$-histogram, sources identified as (a) or (e) have no entry in 
  the black histogram, index and tentative identification from the 
  Supper (1997) catalog (a = foreground star, e = SNR, * = bulge 
  source).
\end{table*}

The SW-sample per definition has no correlation with either a foreground star 
nor a supernova remnant. We define a complementary sample (the C-sample) as
the sample covering a much wider range of candidates fulfilling the 
conditions: 

\begin{equation}
  \begin{array}{l}
  {\rm HR1 < 0.9} \\ 
  {\rm HR2 + \sigma HR2 < -0.1} \\ 
  {\rm exclude\ HR1 + \sigma HR1 < -0.8} 
  \end{array}
\end{equation}

The C-sample comprises 26 objects and is given in Table~1. It turns out to
contain 4 objects correlating with foreground stars and 4 with supernova
remnants. If all identifications are correct then this sample reduces to 18
objects. We introduce quality flags (1=high, 2=medium and 3=low) to qualify 
the overlap of the HR1, HR2 and count rate constraints in the 
$T_{\rm eff}-N_{\rm H}$-plane. ``high'' means that all three contours overlap, 
``medium'' means that the HR1 and the count rate contour overlap and the HR2 
contour overlap within 3-$\sigma$, ``low'' means that the HR1 and the count 
rate contour overlap and the HR2 contour does not overlaps within 3-$\sigma$. 
Especially objects C31 and C33 which have quality flags L and M respectively 
and correlate with SNRs may be discarded. Object C36 shows the characteristics 
of a perfect candidate (all the H1, H2 and CPS contours overlap) and a 
correlation with a SNR may be by chance. C18, C20, C22 and C34 correlate with 
foreground stars but show contour overlap. If they are indeed stars then their
temperatures must be very low (possibly M stars).   

A discussion of the individual candidate sources of the C-sample is beyond the
scope of this article. Interestingly source C26 is located in the bulge of M31
and (if the classification is correct) could harbor a very massive white dwarf
very similar to the SWt transient (cf. Table~1). This source may be recurrent 
or/and very luminous. The latter point is confirmed by the high detected count
rate of $(40\pm1)\ 10^{-3}\ {\rm s^{-1}}$ (cf. the SWt transient has a very 
similar count rate of $(30\pm1)\ 10^{-3}\ {\rm s^{-1}}$).

\section{Estimating the total population}
In the work of DiStefano \& Rappaport (1994) a population of supersoft sources
has been derived e.g. for the M31 galaxy by making certain assumptions about 
their spatial distribution, temperature and luminosity distribution. Here 
the work of DiStefano \& Rappaport can be significantly extended as the sample
derived from the observations has been enlarged significantly and a white 
dwarf mass distribution and a hydrogen column density distribution is derived 
for this sample. 

\subsection{The observation derived $N_{\rm H}$ distribution}
In Figure~1 the $N_{\rm H}$ distribution of supersoft sources from Table~1
with well determined $N_{\rm H}$ values is shown as a histogram. Each source
is distributed with fractional numbers into a number of $N_{\rm H}$ bins 
determined by the uncertainty of the value of $N_{\rm H}$ given in Table~1.
Errors for these fractional numbers have been calculated in the following way.
Distribution of the same fractional numbers in a range of bins which is twice 
as large (twice the error) reduces the fractional number per bin by a factor 
of 2. Therefore we used 0.5 $\times$ the fractional value per bin as the error
per bin. Two histograms are given. The white histogram comprises all sources 
(i.e. 32) for which the $N_{\rm H}$ could be constrained reasonably well, the 
black histogram comprises objects (i.e. 26) not coinciding with foreground 
stars or M31 supernova remnants. The fact that for a comparatively large 
sample of 26 objects hydrogen column densities can be inferred allows to probe
their spatial distribution in the galaxy disk assuming a simple scale height 
law like an exponential law. This method allows not only to probe their 
distribution but also to derive a mean scale height of the detectable 
population. It might well be required to extend this analysis to much larger 
$N_{\rm H}$ values in excess of $5.\times 10^{21}\ {\rm cm^{-2}}$ in order to 
cover the more deeply embedded objects. We extended our calculations to 
absorbing column densities as high as $1.1\ 10^{22}\ {\rm cm^{-2}}$ and 
applied them e.g. to the source with the catalog index~156 (cf. Table~1). We 
find that this source is consistent to be highly absorbed ($N_{\rm H}\sim 
4-10\ 10^{21} {\rm cm^{-2}}$), the total $N_{\rm H}$ column density of the M31
disk at that location is $\sim5.6\ 10^{21} {\rm cm^{-2}}$.

\begin{figure}[htbp]
  \centering{ 
  \vbox{\psfig{figure=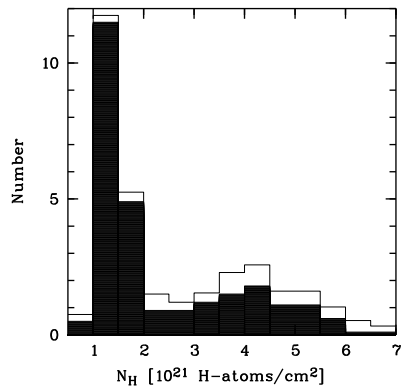,width=6.0cm,%
        bbllx=2.0cm,bblly=1.0cm,bburx=13.0cm,bbury=11.5cm,clip=}}\par
            }
  \caption[]{Distribution of hydrogen absorbing column density for the 
             M31 supersoft sources derived for 32 objects (white histogram)
             and 26 objects (excluding objects identified with foreground
             stars and SNRs, black histogram) from the sample in Table~1.}
\end{figure}

\subsection{The derived white dwarf mass distribution}
Deriving a mass distribution of a galaxy population is by far not trivial.
In case of white dwarfs in supersoft sources it appears to be possible to
derive reliable estimates of the masses using certain assumptions which
have to be shown in later work to be correct or at least not completely
unreasonable (cf. discussion in Kahabka~1998). From the numbers in Table~1 
a mass distribution has been set up (cf. Figure~2). Each source is distributed 
with fractional numbers into a number of $M_{\rm WD}$ bins determined by the 
uncertainty of the value of $M_{\rm WD}$ given in Table~1. The histogram 
comprises all objects (i.e. 26) not coinciding with foreground stars or M31 
supernova remnants for which the mass $M_{\rm WD}$ could be constrained 
reasonably well. The figure shows that only objects with masses in excess 
of $\approx 0.90\ M_{\odot}$ can be detected which is in agreement with the 
expected detection limit in \ros {\sl PSPC} count rates for the used exposure 
time of the observations (cf. Kahabka 1998). The reason is simply that for 
lower white dwarf masses the X-ray luminosities ate too low to be detectable.

\begin{figure}[htbp]
  \centering{ 
  \vbox{\psfig{figure=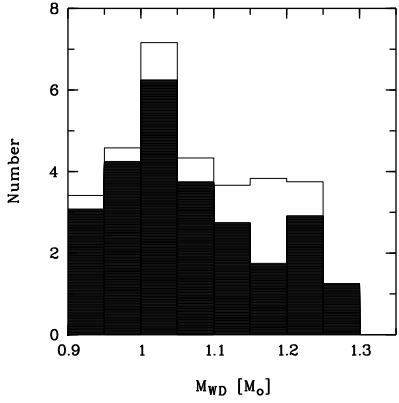,width=6.0cm,%
        bbllx=2.0cm,bblly=1.0cm,bburx=13.0cm,bbury=11.5cm,clip=}}\par
            }
  \caption[]{White dwarf mass distribution of the M31 supersoft sources
             derived for 32 objects (white histogram) and for 26 objects, 
             excluding objects identified with foreground stars and SNRs
             (black histogram). The sample is taken from Table~1.}
\end{figure}

In Figure~3 the cumulative number distribution of white dwarfs in supersoft 
sources with masses $M > 0.5 M_{\odot}$ is given as deduced for the Milky 
Way in the calculations of Yungelson et al. (1996) for the $t_{\rm 3bol}$- 
and the hydrogen-burning shell approximation. $t_{\rm 3bol}$ is the time it 
takes the white dwarf to decline by 3 magnitudes in its bolometric luminosity.
Four distributions of supersoft sources are given for each approximation, the 
distribution of the CV class, the subgiant, symbiotic class and the total 
distribution. The total distributions have been used in our further 
discussion, e.g. to derive from the observed white dwarf distribution the 
predicted total distribution. In the $t_{\rm 3bol}$ approximation 113 (out 
of 1895) objects are expected to be seen in the Milky Way with white dwarf 
masses in excess of $0.90\ M_{\odot}$ after correction for the limited 
visibility due to the $N_{\rm H}$ constraint and about 226 and objects in the 
twice as large M31 galaxy (see section 3.3). In the hydrogen-burning shell 
approximation 107 (out of 1553) objects are expected to be seen in the Milky 
Way with white dwarf masses in excess of $0.90\ M_{\odot}$ after correction 
for the limited visibility due to the $N_{\rm H}$ constraint and about 214 
objects in the twice as large M31 galaxy. If one assumes that one is complete 
for masses in excess of $1.2\ M_{\odot}$ then the observed number of 4 systems 
(cf. Figure~2) would give a total population of $\sim$500 in the 
$t_{\rm 3bol}$ approximation and a total population of $\sim$2000 in the 
hydrogen-burning shell approximation (as 16 for a population of 1895 are 
predicted in the $t_{\rm tbol}$ and 3.3 for a population of 1553 in the 
hydrogen-burning shell approximation). The number of the total population 
deduced from the first approximation appears somewhat small for the M31 galaxy.

\begin{figure}[htbp]
  \centering{ 
  \vbox{\psfig{figure=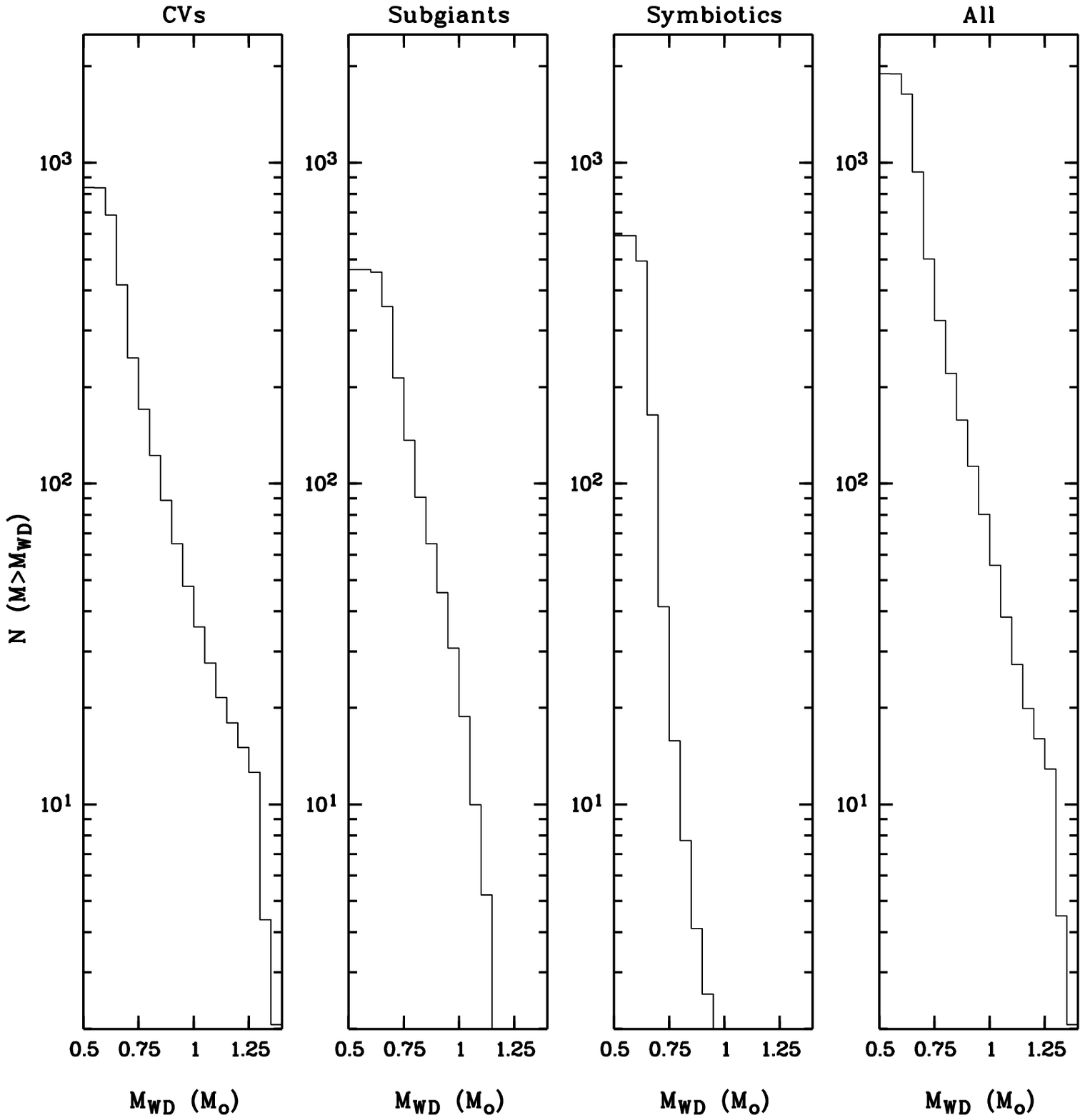,width=8.0cm,%
  bbllx=1.5cm,bblly=2.5cm,bburx=19.0cm,bbury=20.7cm,clip=}}\par
  \vbox{\psfig{figure=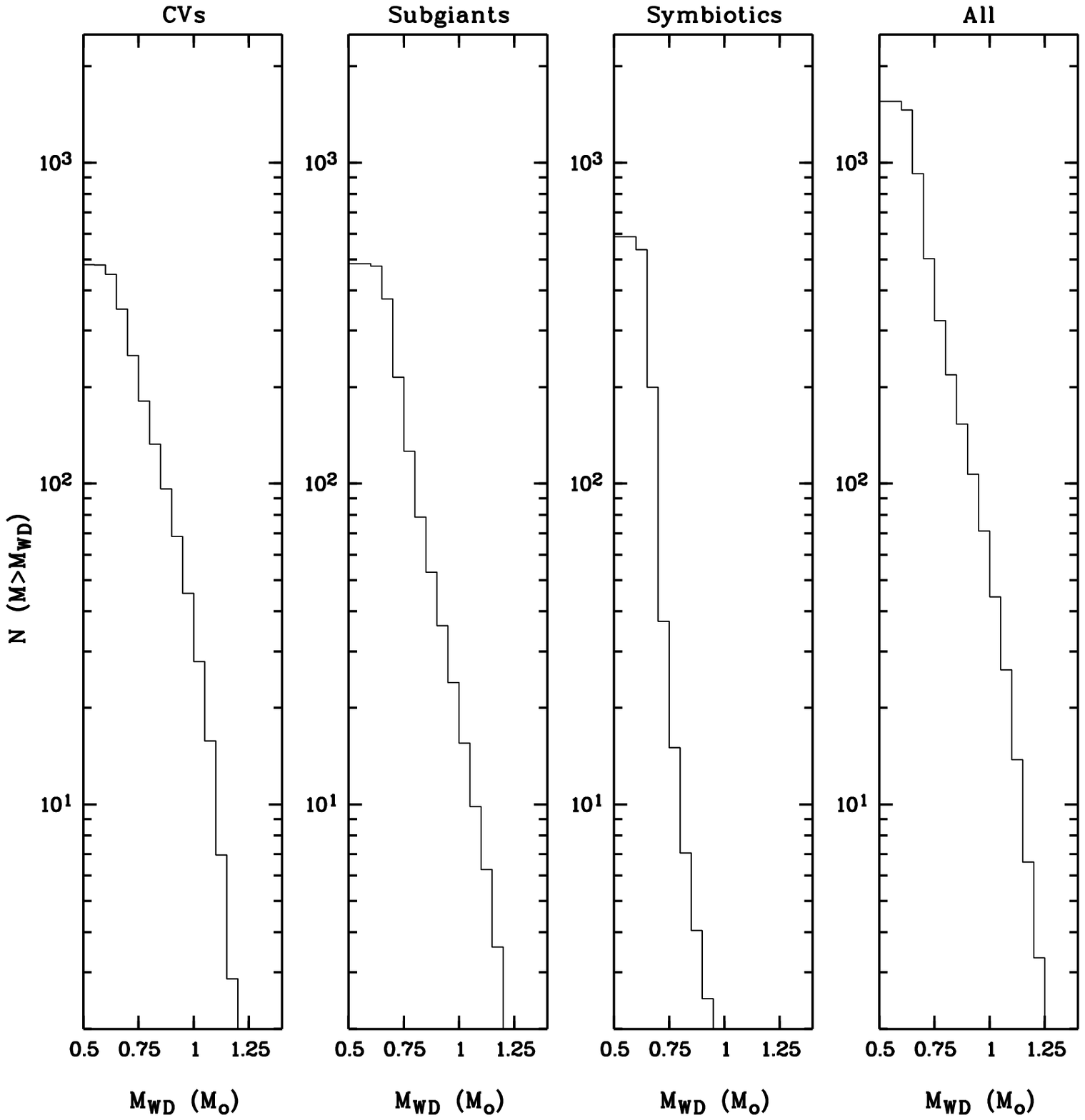,width=8.0cm,%
  bbllx=1.5cm,bblly=2.5cm,bburx=19.0cm,bbury=20.7cm,clip=}}\par
            }
  \caption[]{Cumulative number distribution of CV-type, subgiant,
             symbiotic and total SSS for the Milky Way galaxy. Upper 
             panel: $t_{\rm 3 bol}$ approximation, lower panel: 
             hydrogen-burning shell approximation (deduced from
             Yungelson et al. 1996).}
\end{figure}

\subsection{Correcting for the $N_{\rm H}$ and $M_{\rm WD}$ distribution}
Using the observationally derived $N_{\rm H}$ and $M_{\rm WD}$ distribution 
one can, by comparing with the predicted distributions infer a total number 
of the population. 

In a simple approach a double exponential description as e.g. introduced in
DiStefano \& Rappaport (1994) can be used to describe the source and the 
$N_{\rm H}$ distribution. 
The scale height of the source distribution $h_{\rm s}$ is assumed to be
different from the scale height of the  $N_{\rm H}$ distribution $h_{\rm nh}$.
Expressing the density distribution of the gas and hence the $N_{\rm H}$ 
distribution as

\begin{equation}{\rm
  N_{\rm H} = \int^{\infty}_{z} \rho dz = N_{\rm H}^0 e^{-{\frac{z}{h_{\rm nh}}}},}
\end{equation}

where z is the distance from the galaxy plane, $\rho$ the gas density, and 
$N_{\rm H}^0$ the gas (roughly the hydrogen) column density at the base of the
galaxy disk and expressing the source distribution as

\begin{equation}{\rm
 dN_{\rm s} = -{\frac {N_s^0}{h_s}} e{^{-{\frac {z}{h_s}}}} dz,}
\end{equation}

with $N_{\rm s}^{\rm 0}$ the integrated (total) number of sources in the 
upper hemisphere of the galaxy disk (half of the total population) and 
$h_{\rm s}$ the exponential scale height of the source population, then 
one gets from Equation~5

\begin{equation}{\rm
  z(N_{\rm H}) = -h_{\rm nh} ln (N_{\rm H}(z)/N_{\rm H}^0).}
\end{equation}

Then Equation~6 can be reduced to

\begin{equation}{\rm
  dN_s = N_s^0 {\frac{h_{\rm nh}}{h_s}} 
  (N_H(z)/N_H^0)^{\frac{h_{\rm nh}}{h_s}}
  {\frac{d(N_H/N_H^0)}{(N_H/N_H^0)}}}
\end{equation}

Setting 

\begin{equation}{\rm
  h = {\frac{h_{\rm nh}}{h_{\rm s}}}}
\end{equation}

and 

\begin{equation}{\rm
  n = (N_H/N_H^0)}
\end{equation}

then Equation~8 reduces to

\begin{equation}{\rm
  dN_s = N_s^0 h n^{(h-1)} dn.}
\end{equation}

Equations~8 and 11 give the expected number of sources (above the galaxy disk)
and within  the normalized $N_{\rm H}$ interval $d(N_{\rm H}/N_{\rm H}^0)$. 
From the distribution of observed numbers per $N_{\rm H}$ interval the scale 
height ratio ${\frac{h_{\rm nh}} {h_{\rm s}}}$ can be derived (as well as the 
total number of the population $N_s^0$). This distribution has a powerlaw 
behavior with slope (h-1). If the slope of the source distribution equals the 
slope of the gas distribution then the scale height ratio is 
$h = {\frac{h_{\rm nh}}{\rm h_{\rm s}}} = 1.0$. The scale height for the gas 
may be in the range 150-600~pc for the M31 galaxy (cf. Braun 1991).

\section{Constraining the scale height ratio and the total population
from the normalized $N_{\rm H}$ histogram}
We now will derive from Equation~11 the total population of supersoft sources 
in M31 and the scale height of this population with respect to the scale 
height of the M31 gas. In a first step we define the models used for the M31
gas, in a second step we derive the scaled $N_{\rm H}$ distribution and in a 
third step we apply a least-square fit to the scaled $N_{\rm H}$ distribution.

\subsection{Possible $N_{\rm H}$-models}
From Equation~8 follows that the scaled $n=(N_{\rm H}/N_{\rm H}^0)$ 
distribution has to be considered in order to derive the relative scale height
of the source distribution and the total number of the population. We now 
discuss two possible models for the $N_{\rm H}$-distribution: a schematic one 
by Supper (1997) and a detailled based on radio observations. 

\subsubsection{the Supper-$N_{\rm H}$ model}
In a first approach we use the galaxy $N_{\rm H}$ model given in Figure~12 of 
Supper (1997). The galaxy is divided into 3 concentric ellipsoids covering 
the disk and one circle at the central bulge. The positions of the candidate 
supersoft sources have been projected onto the galaxy disk (cf. Figure~4). 
The disk of M31 may be warped and flaring at the outer part (cf. Evans et al.
1998). 
Such a warping and flaring of the M31 disk may affect the scale height 
assumptions of those supersoft sources which are found in annulus~III of the
$N_{\rm H}$-model of Supper (1997). This point deserves further investigation 
(cf. the galactocentric dependence of a galaxy scale height given by Evans et 
al. 1998 for the M31 disk, cf. also Braun, 1991).

\begin{figure}[htbp]
  \centering{ 
  \vbox{\psfig{figure=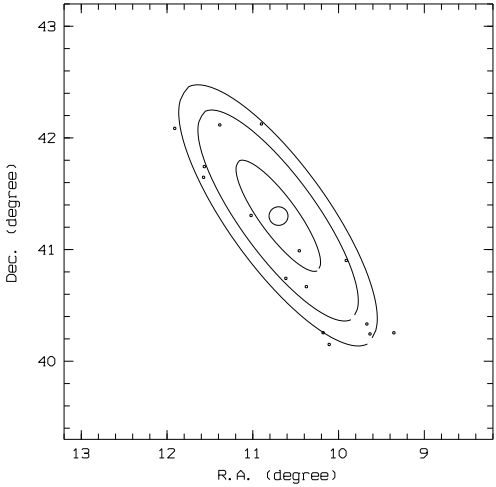,width=7.0cm,%
  bbllx=3.0cm,bblly=2.0cm,bburx=21.0cm,bbury=20.0cm,clip=}}\par
  \vbox{\psfig{figure=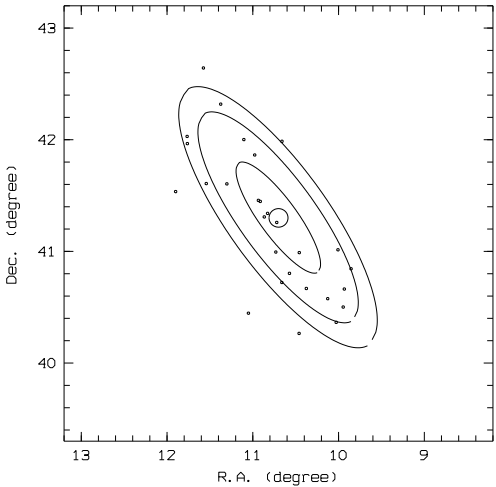,width=7.0cm,%
  bbllx=3.0cm,bblly=2.0cm,bburx=21.0cm,bbury=20.0cm,clip=}}\par
  \vbox{\psfig{figure=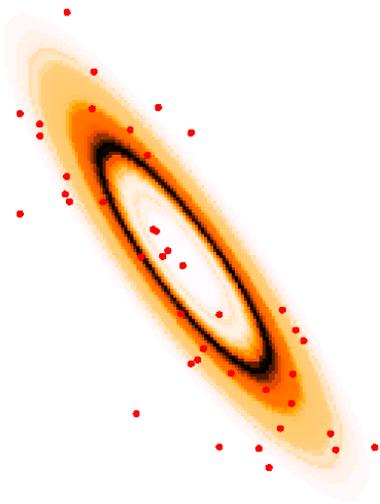,width=7.0cm,%
  bbllx=2.5cm,bblly=10.5cm,bburx=18.5cm,bbury=26.5cm,clip=}}\par
            }
  \caption[]{$N_{\rm H}$ model of M31 as taken from Supper (1997), Figure~12 
             for the SW-sample (upper panel) and C-sample (middle panel).
             Bottom: $N_{\rm H}$ image of M31 deduced from the radial 
             profile of the HI intensity (Urwin 1980). The positions of the 
             supersoft source candidates (the SW-sample and the C-sample) are 
             given (lower panel).}
\end{figure}

\subsubsection{the Urwin-$N_{\rm H}$ model}
As a more refined model for the $N_{\rm H}$ distribution in M31 the Urwin
(1980) model is used. The radial distribution of the hydrogen column is
calculated from the profile given in Figure~8 of Urwin (1980) making use of 
the equation given in Dickey \& Lockman (1990)

\begin{equation}{\rm
  N_H = 1.823\times 10^{18} \int T_b \delta v\ {\rm cm^{-2}}} 
\end{equation}

with the brightness temperature $T_b$, the integral is over the velocity 
profile. With an inclination of $77.5^o$ of the galaxy a maximum column 
density of $1.5\ 10^{22}\ {\rm cm^{-2}}$ is derived for the NE HI~profile 
and a maximum of $8.4\ 10^{21}\ {\rm cm^{-2}}$ for the SW~profile. The 
HI~profile (not corrected for inclination) as determined from Equation~12 
with $\int T_b \delta v$ taken from Figure~8 of Urwin (1980) is shown in 
Figure~5. 

\begin{figure}[htbp]
  \centering{ 
  \vbox{\psfig{figure=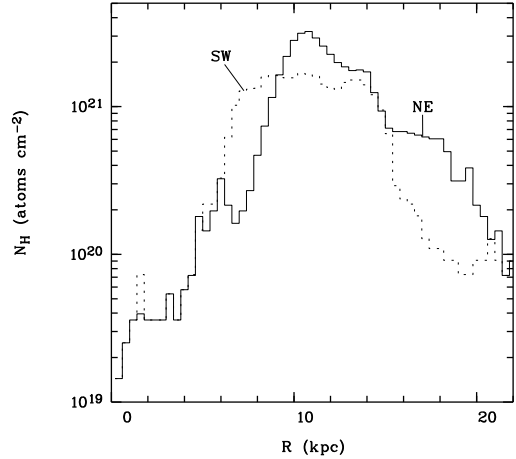,width=7.0cm,%
  bbllx=2.4cm,bblly=1.5cm,bburx=15.5cm,bbury=13.5cm,clip=}}\par
            }
  \caption[]{Radial hydrogen column density profile of M31 not corrected for
             the inclination of the galaxy. The solid histogram gives the 
             north-eastern (NE) profile and dashed histogram gives the 
             south-western (SW) histogram (calculated from the $\int T_b 
             \delta v$ given in Figure~8 of Urwin (1980) and Equation~12).}
\end{figure}

We do not take molecular hydrogen into account. We just mention that a value
of $4\times10^{22}\ {\rm cm^{-2}}$ has been measured at the location RA (1950) 
= $0^h39^m.9$, Decl (1950) = $41^o 14^{'}$ due to molecular hydrogen (cf. 
Urwin 1980, page 257).

\subsection{Expected incompleteness of coverage as a function of hydrogen 
            column density and white dwarf mass}
In Kahabka (1998) the theoretically expected source count rate has been 
derived at the distance of M31 from non-LTE white dwarf atmosphere models 
(model M4) for white dwarf masses in the range $\sim0.95-1.35\ M_{\odot}$ 
under the assumption the source is on Iben's stability line of surface 
hydrogen burning (cf. Iben 1982). We extended these calculations (model M5) 
to white dwarf masses as low as $0.85\ M_{\odot}$ (cf. section~2.1). Taking 
the theoretical white dwarf mass distribution derived by Yungelson (1996) 
into account a number/count~rate diagram was calculated as a function of the 
hydrogen column. The result is given for models M4 and M5 in Figure~6. From 
this diagram the completeness correction factor as a function of the hydrogen 
column has been derived making the following assumption. The X-ray survey
of M31 by Supper (1997) is according to Fig.~13 of Supper complete for {\sl 
ROSAT PSPC} count rates $\approxgt 10^{-3}\ {\rm s^{-1}}$. From our Fig~6 the 
fractional number of sources seen for a specific hydrogen column assuming a 
cut-off count rate of $10^{-3}\ {\rm s^{-1}}$ is derived. This fraction is 
equal to 1.0 for hydrogen columns $\approxlt 8.\ 10^{20}\ {\rm cm^{-2}}$. The 
inverse of this fraction has been used as the correction factor to derive the 
completeness corrected normalized $N_{\rm H}$ histogram making use of the 
specific galaxy $N_{\rm H}$ model. Making e.g. use of the Supper galaxy 
$N_{\rm H}$ model it follows that for all annuli of the galaxy ellipse 
(including the bulge) completeness is not guaranteed and the lower hemisphere 
population is only partially seen. This agrees with the rather small fraction 
of 0.22 of the total galaxy population found for $n>1$ i.e. at the other side 
of the midplane of the disk of M31. From Figure~6 it becomes clear that for 
$N_{\rm H}=5.\times 10^{21}$ only candidates with masses $>1.15 M_{\odot}$ 
are detectable. This means adding the mean foreground $N_{\rm H}$ of 
$6.\times 10^{20}$ in Supper's model annulus III is opaque for the lower 
hemisphere population with $M_{\rm WD}\le 1.15 M_{\odot}$ but the bulge and 
annulus II are transparent for somewhat less massive white dwarfs 
($M_{\rm WD}\le 1.05 M_{\odot}$).

\begin{figure}[htbp]
  \centering{ 
  \vbox{\psfig{figure=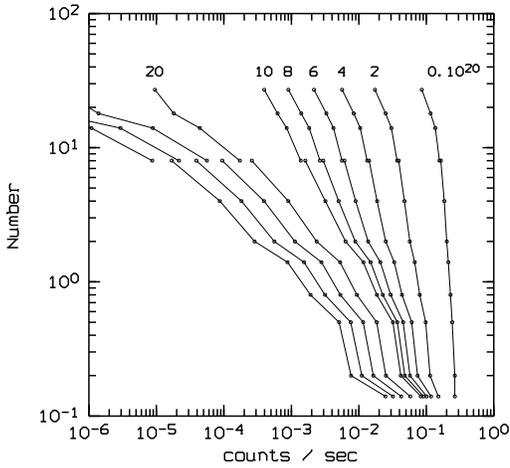,width=7.0cm,%
  bbllx=2.0cm,bblly=1.6cm,bburx=13.7cm,bbury=12.3cm,clip=}}\par
            }
  \caption[]{Theoretical number/count rate diagram of supersoft sources for a 
             galaxy of the size of the Milky Way and for a distance of 700~kpc
             (M31). Note that a galaxy like M31 has a mass twice as large as 
             the Milky Way. These numbers follow from the population synthesis
             calculations of Yungelson et al. (1996) for the Milky Way galaxy.
             Labels mark hydrogen column densities ($10^{20}\ {\rm cm^{-2}}$).
             Dots mark white dwarf masses (counted from bottom to top) 
             1.34,1.29,1.25,1.20,1.15,1.10,1.00(twice),0.95,0.90,0.85 
             $M_{\odot}$. This figure shows the results for model M5
             ($M_{\rm WD}\le 1.0 M_{\odot}$) and for model M4
             ($M_{\rm WD}\ge 1.0 M_{\odot}$).}
\end{figure}

\begin{figure}[htbp]
  \centering{ 
  \vbox{\psfig{figure=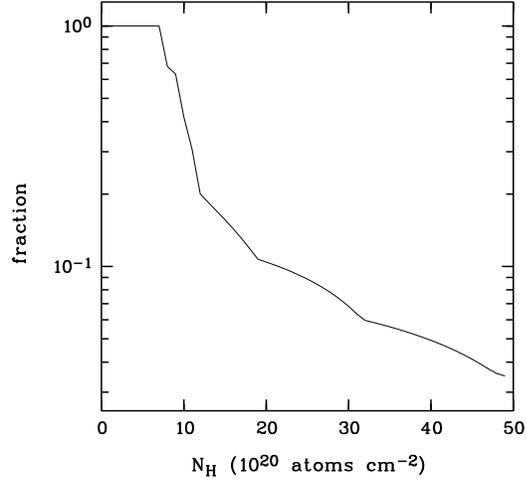,width=7.0cm,%
  bbllx=4.0cm,bblly=1.5cm,bburx=15.5cm,bbury=12.5cm,clip=}}\par
            }
  \caption[]{Fraction of sources with white dwarf masses above
             $0.90\ M_{\odot}$ seen in the Andromeda galaxy (M31) for
             different hydrogen absorbing columns. This fraction is equal
             to 1.0 for columns below $8.\ 10^{20}\ {\rm cm^{-2}}$, which 
             means completeness is fulfilled and decreases for increasing 
             columns.}
\end{figure}

Making use of the number/count rate distribution derived from model M4 and 
M5 the fraction of objects seen for different hydrogen columns has been 
calculated and the result is given in Figure~7. This fraction is equal to 
1.0 for columns below $8. 10^{20}\ {\rm cm^{-2}}$, which means completeness is 
fulfilled, and decreases for increasing columns.

\subsection{Number of supersoft sources expected for the Supper-$N_{\rm H}$
model}
The scaled $n=(N_{\rm H}/N_{\rm H}^0)$ values have been derived making use of 
the $N_{\rm H}$ ranges given in Table~1, deriving the local $N_{\rm H}^0$ 
values from the Supper $N_{\rm H}$ model and by applying the completeness 
correction. The scaled $N_{\rm H}$ histogram is plotted in Figure~8. This 
distribution extends from n=0 to n=2 and comprises both hemispheres of the 
galaxy. It turns out that of the 18 sources in the distribution 22\% (4 of 18)
fall below the galaxy disk. This may not be unexpected as the masses and hence
temperatures of the white dwarfs involved are substantial and the total 
$N_{\rm H}$ of the galaxy disk is only in one $N_{\rm H}$ ellipse large enough
(i.e. $\approx 7.7\times 10^{21} {\rm cm^{-2}}$) to hide the lower hemisphere 
population nearly completely. 

\begin{figure}[htbp]
  \centering{
  \vbox{\psfig{figure=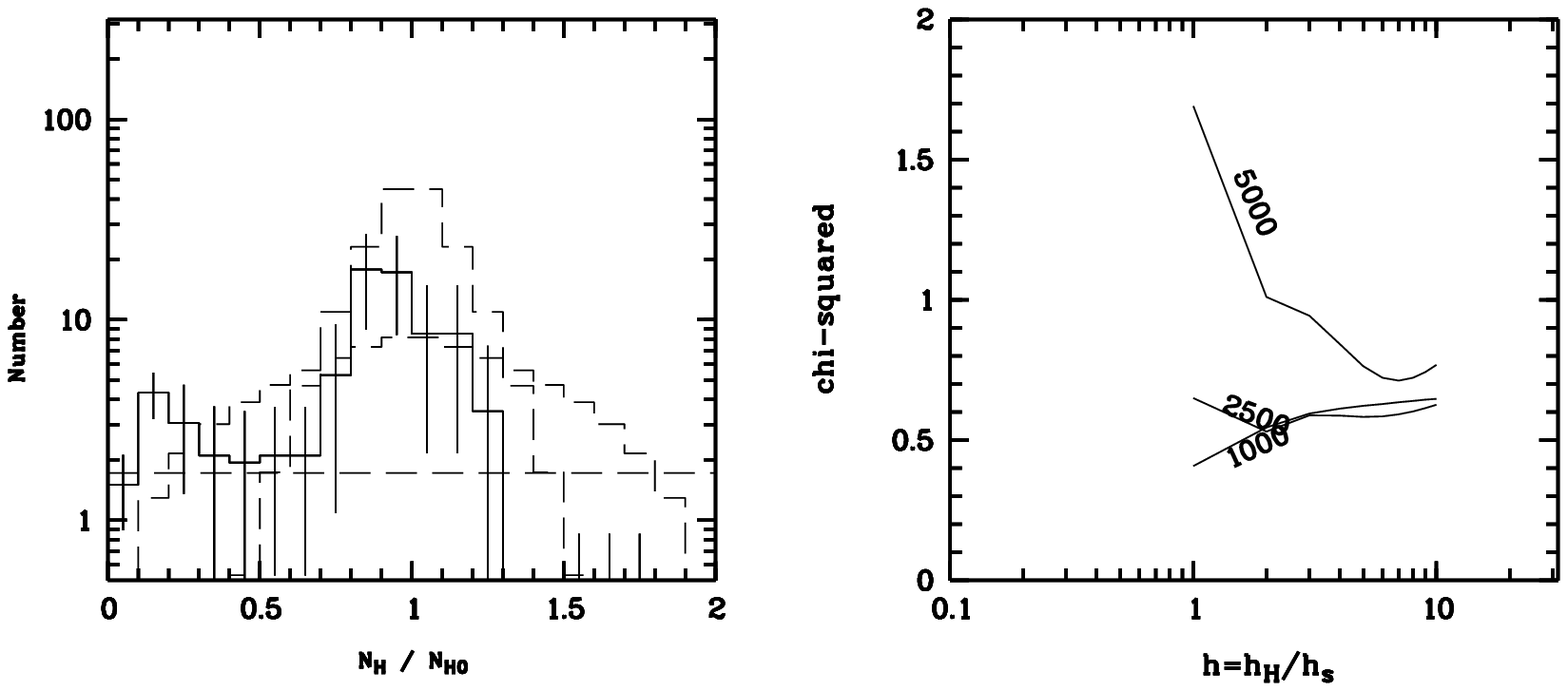,width=7.0cm,%
  bbllx=1.2cm,bblly=1.7cm,bburx=9.5cm,bbury=9.5cm,clip=}}\par
            }
  \caption[]{Normalized $N_{\rm H}/N_{\rm H}^{\rm 0}$ distribution corrected 
             for completeness with 80 (corrected) candidates with well 
             determined $N_{\rm H}$ values in the n=0.0-1.0 interval. The 
             best-fit for a population of 1000, 2500, and 5000 sources is 
             given as dashed histogram.}
\end{figure}

\begin{figure}[htbp]
  \centering{ 
  \vbox{\psfig{figure=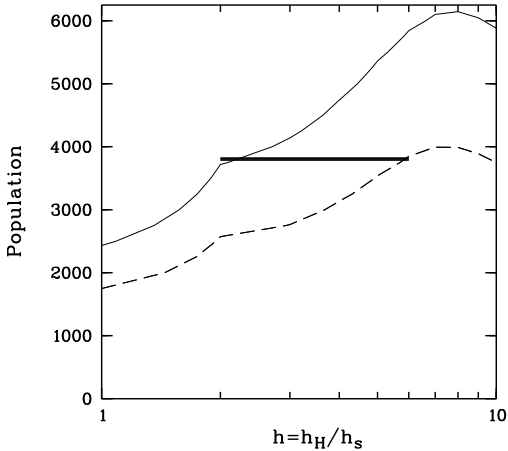,width=7.0cm,%
  bbllx=2.5cm,bblly=1.5cm,bburx=14.5cm,bbury=12.2cm,clip=}}\par
            }
  \caption[]{Result of chi-squared fit (of Equation~11) to the distribution 
             of Figure~8. The scale height ratio $h=h_H/h_s$-population plane 
             is shown. The fit has been applied to the $0<n<1.0$ distribution 
             (upper hemisphere). The 90\% and 95\% confidence bounds are shown
             as dashed and solid lines. The bar gives the estimate derived 
             from the galactic population, cf. section~4.4}
\end{figure}

The total population can be inferred with Equation~11. Using the number of 
the corrected population above and below the galaxy disk of 80 (which is 
uncertain in the range 30-130 considering the errors, cf. Figure~8) and 
assuming that only a fraction of 6.0\% of the whole population in the 
$t_{\rm 3bol}$ and of 6.9\% in the hydrogen-burning shell approximation is 
covered as only objects with white dwarf masses in excess of 0.90$M_{\odot}$ 
are detected a total population of 1300 (500-2200) and 1200 (430-1900) 
respectively is derived for the Andromeda galaxy. These numbers are consistent
with the range of $\sim$800-5000 supersoft sources predicted from the 
population synthesis calculations of DiStefano \& Rappaport (1994). The 
distribution would be consistent to be centered at $N_{\rm H}/N_{\rm H}^{\rm 
0} = 1$. This fits with a disk population of a scale height 
significantly smaller than the gas scale height. A scale height ratio can be 
constrained from this histogram. This means the scale height of the source 
distribution $h_s$ can be determined with Equation~9 if the scale height of 
the gas distribution $h_{\rm nh}$ is known. As a function of galactocentric 
radius $h_{\rm nh}$ varies from 150~pc to 600~pc (Braun 1991). A chi-squared 
fit has been applied to the normalized $N_{\rm H}$ distribution. The result 
of a chi-squared fit of Equation~11 to the distribution given in Figure~8 is 
given in Figure~9. The range of the population follows from the chi-squared 
fit to the measured distribution taking the errors into account. A total 
population of 1,800-5,800 sources is obtained for h-values 1$<$h$<$6, which 
means for a source population which is more confined to the galaxy disk than 
the gas distribution. If there is a large population of supersoft sources in 
M31 then the sources are very confined to the galaxy plane. There may exist a 
number of the order 200 hot ($>10^5\ K$) and X-ray luminous planetary nebula 
nuclei in a spiral galaxy of the size of M31 according to the estimates of 
Iben \& Tutukov (1985). They can be a minor sub-population of a larger 
population of luminous supersoft sources but with a larger scale height 
(h$<$1). From the formal fit of Equation~11 to the normalized $N_{\rm H}$ 
distribution (making use of the Supper-$N_{\rm H}$ model) we would exclude 
that a population of hot and luminous planetary nebula nuclei (of order of 200
objects) alone account for the observed sample.

\subsection{Number of sources expected with the Urwin-$N_{\rm H}$ model}
The normalized $N_{\rm H}$ histogram is also calculated by making use of 
the Urwin $N_{\rm H}$ model for the north-eastern (NE) part of the galaxy. 
This $N_{\rm H}$ model consists of a radial distribution with 57 rings. This 
distribution has been converted into a galaxy $N_{\rm H}$ model of M31 by 
assuming an inclination of the galaxy of $77.5^o$ (cf. Figure~4). This is a 
more refined model than the Supper-$N_{\rm H}$ model. The normalized 
distribution is given in Figure~10. This distribution appears to cover
only parts of the normalized $N_{\rm H}$-bins. The main histogram extends 
over the range ${\frac{N_{\rm H}}{{N_{\rm H}}_0}}$=0.0-0.6. This fact can be 
explained if one considers the galactocentric distribution (12-16~kpc) of 
the sources which fall into this interval (cf. section~5) and the projected 
hydrogen columns of the M31 galaxy $\sim(4-9)\ 10^{21}\ {\rm cm^{-2}}$ for 
these radii. The hydrogen columns are that large that indeed only part of the 
upper hemisphere population is detectable in agreement with the histogram 
extending to values well below n=1.0. The entries in the histogram for 
n$\sim1.5-2.0$ are from the population found at radii 18-23~kpc. Here the 
projected hydrogen columns of $\sim(1-3)\ 10^{21}\ {\rm cm^{-2}}$ are lower 
and the lower hemisphere population is detectable. But this part of the 
histogram is not very significant. We constrain our fit of Equation~11 only 
to the n=0.0-0.6 regime. This allows to constrain the size of the population 
and the scale height ratio. As we do not cover the top of the distribution we 
are not able to determine an upper bound for the population. Only by 
constraining the scale height ratio to realistic values for stellar 
populations we can determine an upper bound for the population.

\begin{figure}[htbp]
  \centering{
  \vbox{\psfig{figure=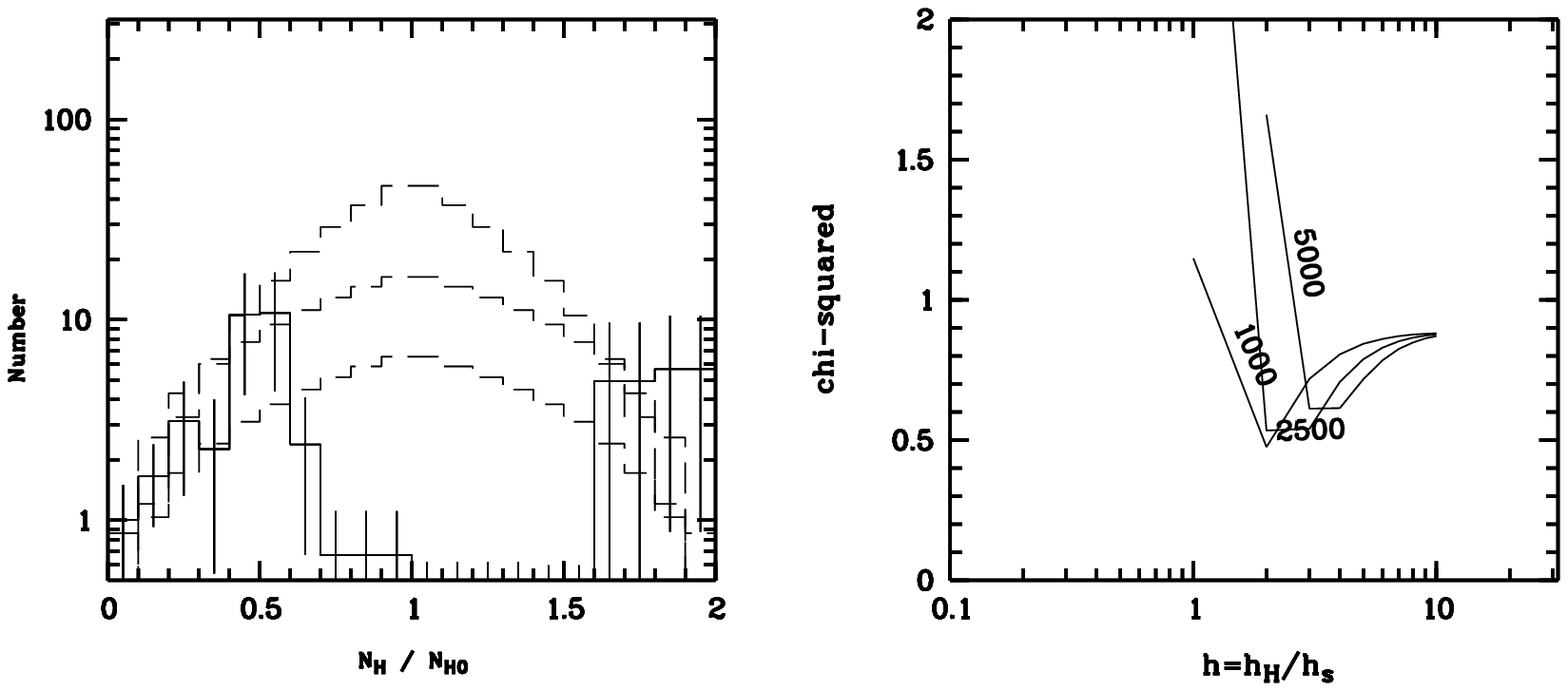,width=7.0cm,%
  bbllx=1.2cm,bblly=1.7cm,bburx=9.5cm,bbury=9.5cm,clip=}}\par
            }
  \caption[]{Normalized $N_{\rm H}/N_{\rm H}^{\rm 0}$ distribution corrected 
             for completeness with 24 (corrected) candidates with well 
             determined $N_{\rm H}$ values in the n=0.0-0.6 interval. 
             The best-fit for a population of 1000, 2500, and 5000 sources
             is given as drawn and dashed histograms.}
\end{figure}

\begin{figure}[htbp]
  \centering{ 
  \vbox{\psfig{figure=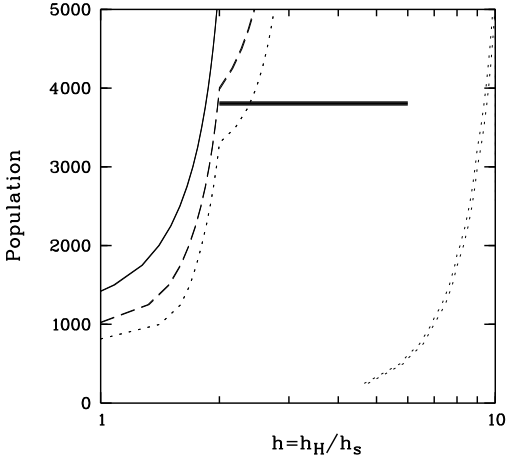,width=7.0cm,%
  bbllx=2.5cm,bblly=1.5cm,bburx=14.5cm,bbury=12.2cm,clip=}}\par
            }
  \caption[]{Result of chi-squared fit (of Equation~11) to the distribution 
             of Figure~10. The confidence plane for the scale height ratio 
             $h=h_H/h_s$-population plane shown. The 95\% confidence bound is 
             given as solid line and the 90\% confidence bound as dashed line.
             The 10\% confidence is given by a dotted line. The fit has been 
             applied to the $0<n<0.6$ distribution (part of upper hemisphere).}
\end{figure}

The size of the population as derived with the Urwin-$N_{\rm H}$ model is
consistent to be in the range $\sim$1000-10,000 sources for a scale height 
ratio h=1-5 (cf. Figure~11).

As a refined model the radial profiles of the north-eastern (NE) and the
south-western (SW) galaxy as given in Urwin (1980) have been used to
calculate a $N_{\rm H}$-map of the galaxy and to deduce the hydrogen-column
at the location of each supersoft source. The normalized $N_{\rm H}$ 
distribution has been calculated which is given in Figure~12. This 
distribution extends over a similar $N_{\rm H}/N_{\rm H0}$ range as for the
Urwin model. A fit of Equation~11 to this distribution for a population is 
given in Figure~13 as a function of the scale height ratio $h = h_{\rm H} / 
h_{\rm s}$. This distribution again extends mainly over the n=0.0-0.6 interval
(see discussion above). The size of the population is $\sim$1000-10,000 for 
a scale height ratio h=1-5. There are sources from Table~1 which fall beyond 
the n=2 limit and are rejected (in the specific $N_{\rm H}$-model). For the 
NE-SW Urwin model these sources are found either at radii $>$15 kpc or at 
radii $<$5 kpc. The nature of these sources is unclear or the 
$N_{\rm H}$-model is still too crude (but see discussion in section~5). Some 
sources correlate with a foreground star or a M31 supernova remnant. Another 
possibility is that these sources are located at a large distance from the 
galaxy plane ($>$500~pc) and are projected due to the considerable inclination
of the galaxy towards the wrong reference hydrogen column. But this appears 
to be quite unlikely.

\begin{figure}[htbp]
  \centering{ 
  \vbox{\psfig{figure=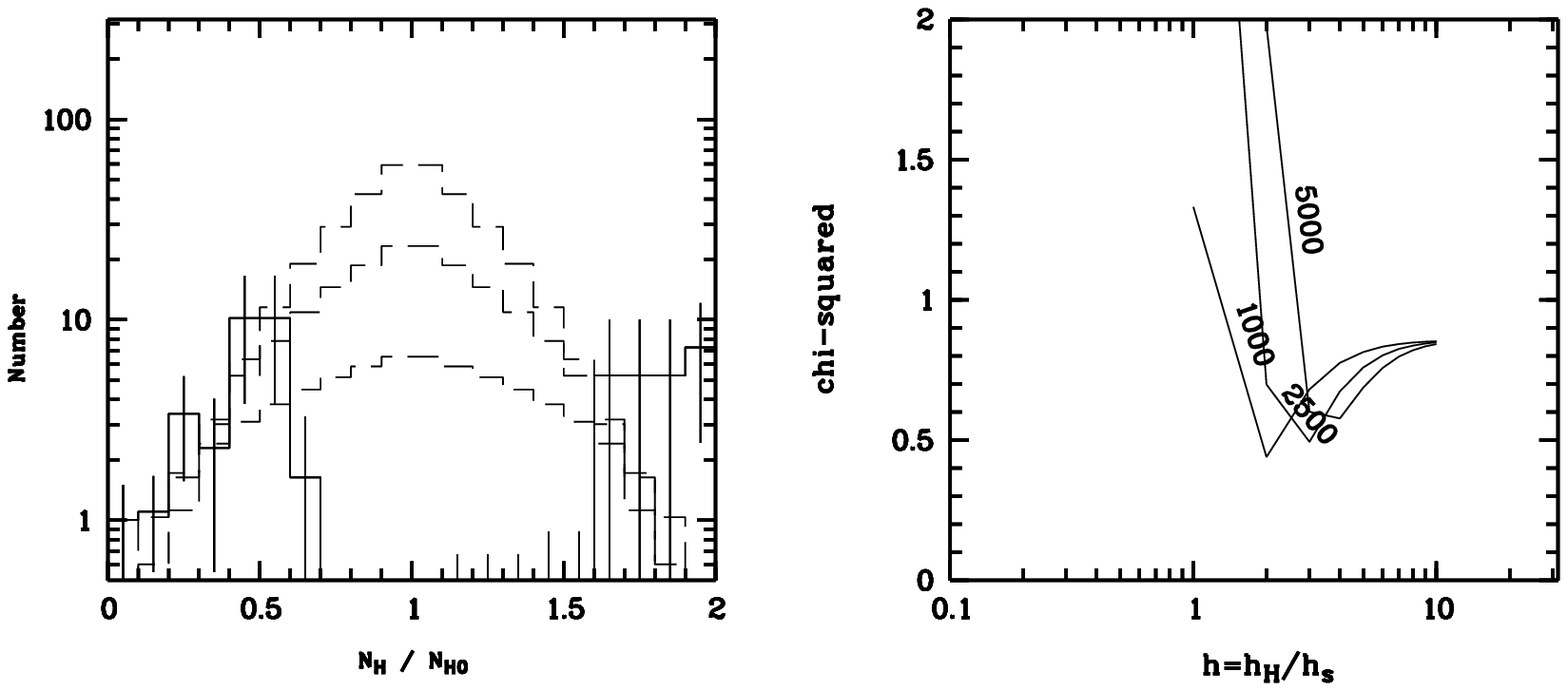,width=7.0cm,%
  bbllx=1.3cm,bblly=1.7cm,bburx=9.5cm,bbury=9.2cm,clip=}}\par
            }
  \caption[]{Normalized $N_{\rm H}/N_{\rm H}^{\rm 0}$ distribution corrected
             for completeness with 23 (corrected) candidates with well 
             determined $N_{\rm H}$ values deduced with the Urwin $N_{\rm H}$ 
             model (NE and SW galaxy) in the 
             $n={\frac{N_{\rm H}}{{N_{\rm H}}_0}}$ interval n=0.0-0.6. The 
             dashed histograms give the best-fit for a population of 1000, 
             2500, and 5000 sources.}
\end{figure}

\begin{figure}[htbp]
  \centering{ 
  \vbox{\psfig{figure=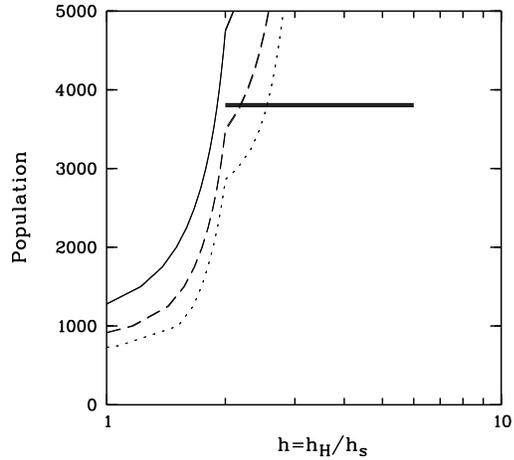,width=7.0cm,%
  bbllx=2.5cm,bblly=1.5cm,bburx=14.5cm,bbury=12.2cm,clip=}}\par
            }
  \caption[]{Result of chi-squared fit (of Equation~11) to the distribution 
             of Figure~12. The scale height ratio $h=h_H/h_s$-population 
             plane is shown. The 90\% and 90\% confidence bounds are shown
             as dashed and solid lines. The 10\% confidence bound is shown
             as dotted line. The fit has been applied to $0<n<0.6$ 
             distribution (part of upper hemisphere). The bar gives the 
             estimate derived from the galactic population, cf. section~4.3.}
\end{figure}

The NE-SW model may describe the distribution of the hydrogen column in M31
in a good approximation. It becomes evident that in the range of galactocentric
radii 12-16~kpc where most supersoft sources are found the hydrogen column is
that large that only part of the upper hemisphere population is visible. The
total population can be constrained dependent on the scale height ratio. 

In Table~2 the size of the population of supersoft sources in M31 as derived 
from different galaxy $N_{\rm H}$ models is summarized. In the Supper 
$N_{\rm H}$ model numbers have been derived from the n=0-1 histogram (the 
complete upper galaxy hemisphere) and in the Urwin $N_{\rm H}$ model from the 
n=0-0.6 histogram (60\% of the upper galaxy hemisphere). Interestingly the 
range of the population derived from different $N_{\rm H}$ models does not 
differ much. This may be due to the fact that the errors associated with the 
(corrected) numbers are substantial due to the small number of selected 
sources. In order to better confine the range of the population detections of 
supersoft sources in the 12-16~kpc ring for values n$>$0.6 are required. Such 
sources are heavily absorbed, they have hydrogen columns  $N_{\rm H}>2.5-5\ 
10^{21}\ {\rm cm^{-2}}$ according to Figure~6 and they are only detected in 
the {\sl ROSAT} 1991 survey of Supper if the white dwarf mass is in excess 
of $1.2\ M_{\odot}$ (see possible candidates in the C-sample, cf. Table~1).

\begin{table}
  \caption[]{Population of supersoft sources in M31 derived from a
             chi-square fit of Equation~11 to the normalized $N_{\rm H}$ 
             histogram for different galaxy $N_{\rm H}$ models and different
             scale height ratios $h=\frac{h_{\rm nh}}{{h_{\rm s}}}$.}
  \begin{flushleft}
  \begin{tabular}{lcccc}
  \hline
  \hline
  \noalign{\smallskip}
                               &\multicolumn{3}{c}{population/1000}       \\
  $N_{\rm H}$ model & fit applied to &  h=1    &   2      &   5           \\
  \noalign{\smallskip}
  \hline
  \noalign{\smallskip}
  Supper      & (0$<$n$<$1.0)  & 1.8-2.5 & 2.5-3.8   & 3.5-5.3            \\
  Urwin       & (0$<$n$<$0.6)  & $<$1.5  & $<$5      & $<$15              \\
  Urwin-NE-SW & (0$<$n$<$0.6)  & 0.5-1.3 &   2-5     &  5-15              \\
  \noalign{\smallskip}
  \hline
  \noalign{\smallskip}
  \end{tabular}
  \end{flushleft}
\end{table}

\subsection{Comparison with the galactic population}
There is evidence that the group of observed galactic supersoft sources is
larger than assumed. Patterson (1998) proposes three sources to belong to
this family, e.g. V~Sge, T~Pyx and (possibly) WX~Cen. These blue and optically
bright binary systems have orbital periods of 12, 2, and 10 hours. Assuming
distances of 1.3, 2.5, and 1.4 kpc the objects are found 200, 430, and 16 pc
above the galactic plane. The two ``standard'' galactic supersoft sources
RX~J0925.7-4758 and RX~J0019.8+2156 are 33 and 840 pc above the galactic
plane assuming distances of 1~kpc. Assuming an exponential z-distribution
(cf. Equation~6) and assuming a scale height for the source population 
$h_{\rm s}$ the total population can be constrained in order to be 
consistent with this sample. This is an independent consistency check for
the distribution and size of the galactic sample. Assuming a scale height of
150~pc the population has to be greater than 270 in order to explain the
discovery of one source such as RX~J0019.8+2156 at such a large scale height.
Assuming a much smaller scale height of 30~pc the probability of observing
one RX~J0019.8+2156 is negligible. The scale height of a population of
supersoft sources which can explain RX~J0019.8+2156 has to be larger than
105~pc if the total population is $<$3000. This is not a problem as 105~pc
is still a small scale height for stellar populations. T~Pyx is a recurrent
supersoft source which is found 200~pc above the galactic plane. T~Pyx may
harbor a massive white dwarf as it has a recurrence period of 20~years.
According to the population synthesis calculations of Yungelson (1996) there
may be 113 galactic supersoft sources which are more massive than $0.9\ 
M_{\odot}$. These sources can become recurrent supersoft sources with such
a recurrence period (cf. Kahabka 1995). Assuming a scale height of 105~pc 
for the source distribution 2 sources are expected to be found at a distance 
from the galactic plane as large as in T~Pyx (430~pc). To observe one T~Pyx
is therefore in full agreement with this number. The distance from the
galactic plane of all other supersoft sources is considerably smaller and
is in agreement with such a population. The conclusion is that the prediction
of a population of $\sim$1900 supersoft sources in the Milky Way by the
population synthesis calculations of Yungelson (1996) is in agreement with
the so far discovered galactic population if the scale height is $\sim$100~pc.
One expects then 0.6 systems to be observed at the distance from the galactic
plane of 840~pc, the distance of RX~J0019.8+2156.

Assuming a scale height for the galactic supersoft sources of 100~pc and a
scale height for the gas of 200-600~pc a value $h=2-6$ (Equation~9) is
derived. Scaling with the mass ratio of the Andromeda galaxy and the Milky
Way galaxy, which is about 2, one expects from the population synthesis 
calculations of Yungelson (1996) that there exists a population of $\sim$3800 
supersoft sources in M31. Such a population having $h=2-6$ is consistent with
the chi-squared fit to the normalized $N_{\rm H}$ distribution given in
Figure~9, 11 and 13. There is still the possibility of a bi-modal population
consisting of a more extended population (e.g. the CV-type supersoft sources)
and a more to the galaxy plane confined population (e.g. the subgiant class).
We fitted such a bi-modal population to the normalized $N_{\rm H}$ histogram
of Figure~12 and find that an extended ($3>h>1$) population of 500 sources and
a confined (h$<$10) population of $\le 6000$ sources is possible.

\section{The spatial distribution}
The 26 supersoft sources found in M31 are distributed over the whole galaxy 
disk (cf. Figure~4 for the distribution of the SW and the C-sample, cf. also
Figure~14 and Figure~15). This may be in favor for a disk population. Recent 
spatial studies of novae in M31 using a Monte Carlo simulation have been 
performed by Hatano et al. (1997). From this study it follows that the ratio 
of bulge to disk population is about 1/2 similar to the ratio of bulge to disk
mass of this galaxy although there is some controversy about this subject. In 
comparison for the Milky Way a ratio of 1/7 is found both for the novae and 
the mass. The M31 bulge has in the model of Hatano a radius of $\sim$6~kpc 
(equivalent to a projected size of 0.5$^{o}$). A more extended discussion on 
the distribution of novae in M31 can e.g. be found in Capaccioli et al (1989), 
Rosino et al. (1998) and Yungelson et al. (1997). Interestingly there are 6 
of the 31 (likely) candidate supersoft sources found in the bulge region. But 
three of these objects correlate eiher with a galactic foreground star or a  
M31 supernova remnant. A ratio 1/(5-10) is derived for supersoft sources 
detected in the bulge compared to objects detected in the disk. This ratio 
reduces to 1/(4-7) if only ``accepted'' systems are considered (cf. 
Figure~15). There may be a chaining of supersoft sources (from the C-sample) 
along a $\sim$(12-16)~kpc arm (cf. Figure~4,5, 14 and 15). This feature may 
also be found in the model of Hatano (cf. his Figure~3). The spatial 
distribution of the SW-sample (with a mean radius of 14~kpc) is consistent 
with a 12-16~kpc spiral arm of the M31 galaxy (cf. Figure~6 of Braun 1991). 
Another grouping of supersoft sources in a 18-21~kpc ring (cf. Figure~15) may 
be connected to a 18-24~kpc spiral arm (cf. Braun 1991). There are no 
detections within the 6-12~kpc ring (the only exception may be the source with
the catalog index 212). In this ring the (projected) hydrogen column reaches 
values up to $1.5\times10^{22}\ {\rm cm^{-2}}$.

\begin{figure}[htbp]
  \centering{ 
  \vbox{\psfig{figure=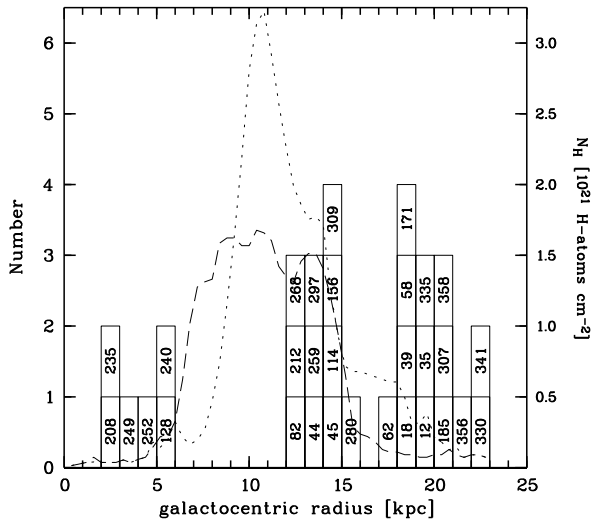,width=8.00cm,%
  bbllx=3.7cm,bblly=1.7cm,bburx=16.7cm,bbury=13.2cm,clip=}}\par
            }
  \caption[]{Distribution of supersoft sources in M31 (solid histogram) 
             as a function of the galactocentric distance in M31. 
             Also shown is the radial distribution of the hydrogen column 
             density for the NE Urwin model (dotted line) and for the SW 
             Urwin model (dashed line) which is not corrected for the 
             inclination of the galaxy. The catalog indices of individual 
             sources are given (cf. Table~1).}
\end{figure}

In Figure~15 we also show the galactocentric distribution of the M31 novae
(from Sharov \& Alksnis 1991, 1992), blue stars (B-V$<$0.3) and Cepheids 
(from Magnier et al. 1992, 1997 and Haiman et al. 1994). Novae belong to an
old stellar population and are preferentially detected in the bulge (at
galactocentric radii $\approxlt$6~kpc). Blue stars belong to a young stellar 
population and trace the galaxy light. Cepheids belong to a somewhat older 
population. The distribution of Cepheids trace the spiral arms of M31 where 
recent star formation is taking place. They are found (in the distribution) 
predominantly within 8-15~kpc. For a consideration of the completeness of the 
Cepheid sample see Magnier et al. (1997). Actually there could be a second 
Cepheid peak at galactocentric radii 18-22~kpc where another M31 spiral arm 
is found and where a peak in the supersoft distribution is found. But the 
Magnier survey apparently did not cover this region.

\begin{figure}[h]
  \centering{ 
  \vbox{\psfig{figure=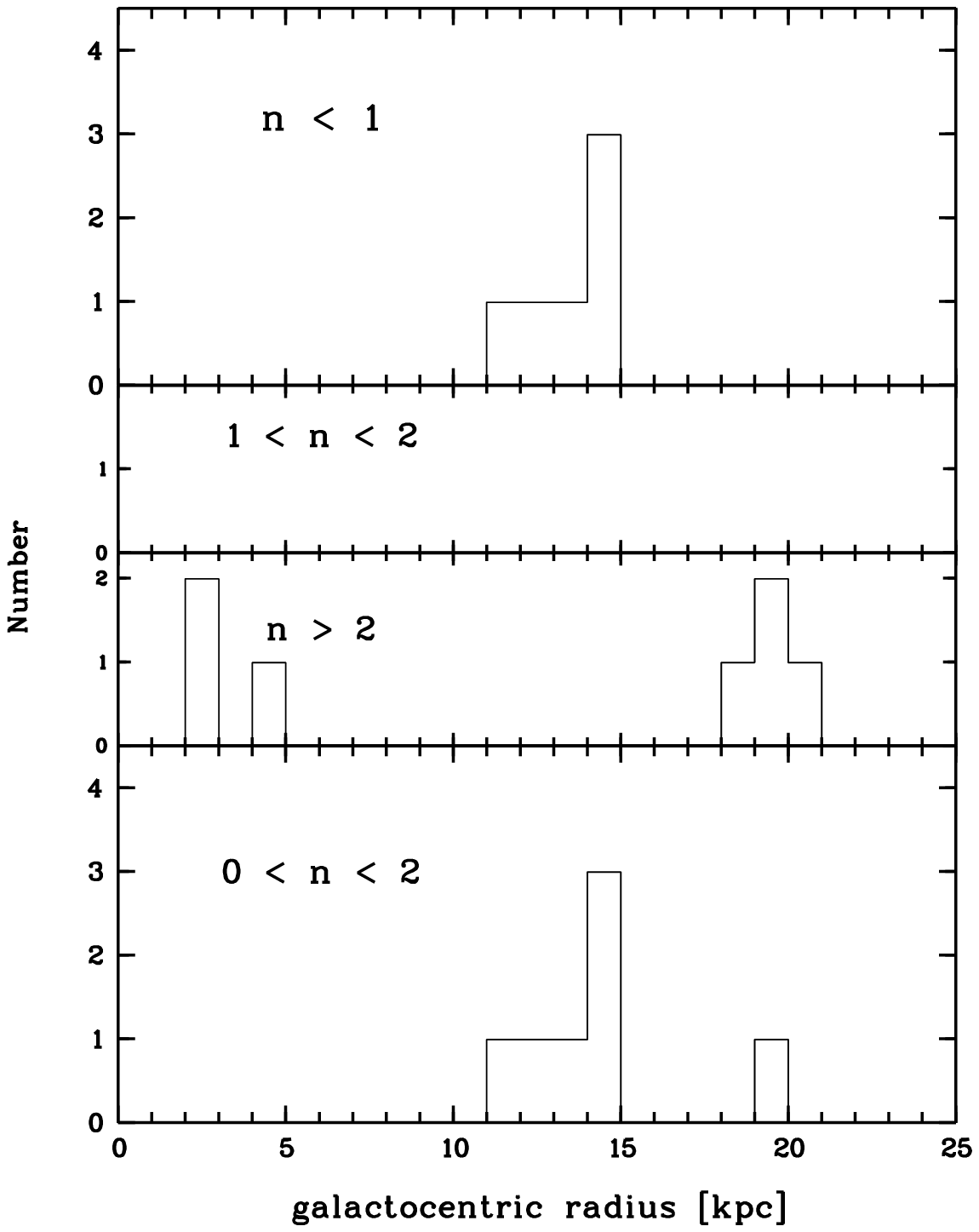,width=6.0cm,%
  bbllx=4.5cm,bblly=1.5cm,bburx=16.5cm,bbury=16.5cm,clip=}}\par
  \vbox{\psfig{figure=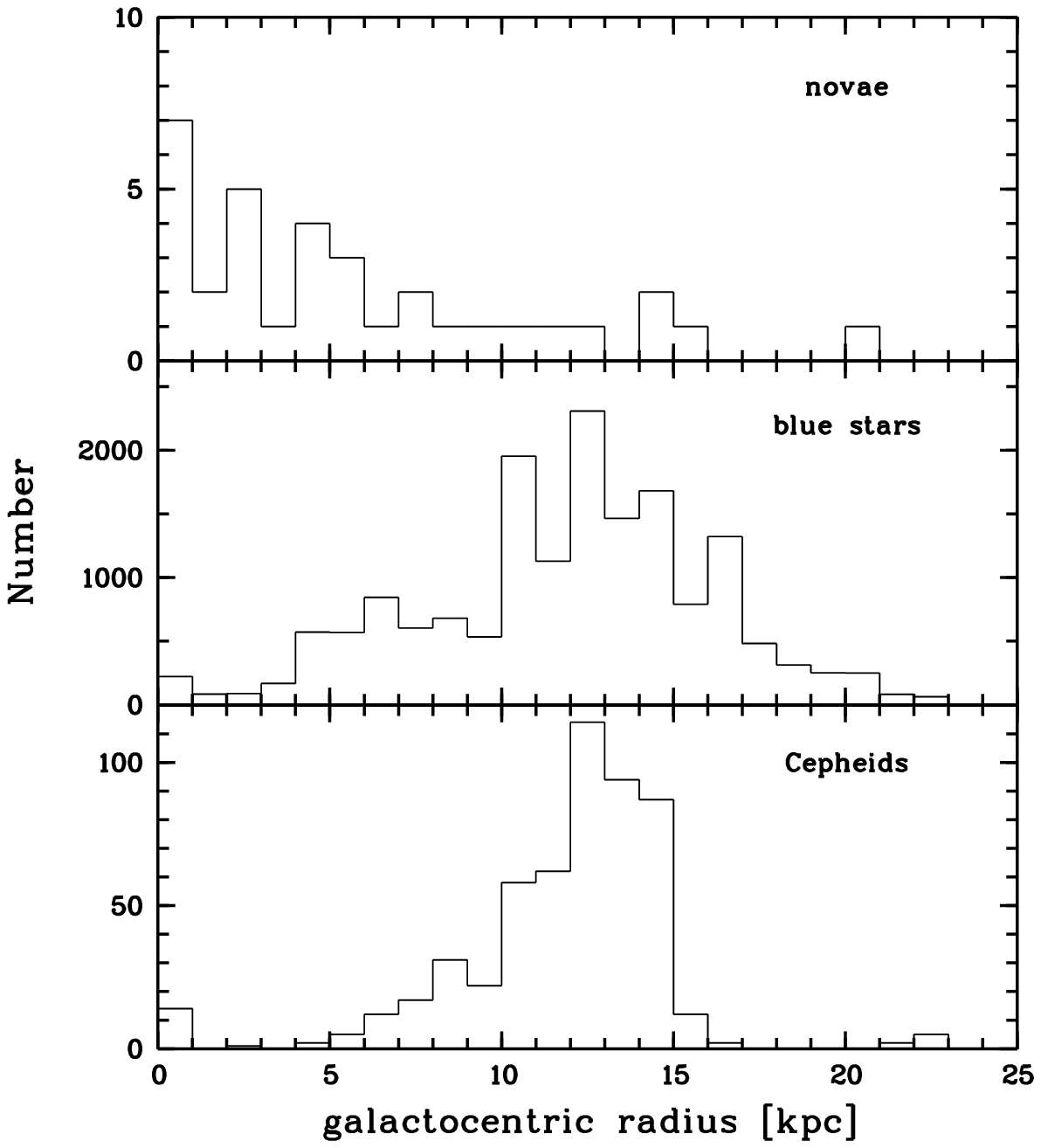,width=6.0cm,%
  bbllx=4.2cm,bblly=1.5cm,bburx=16.2cm,bbury=15.5cm,clip=}}\par
            }
  \caption[]{Upper panel: galactocentric distribution of supersoft sources 
             in M31. Separate histograms are shown for objects not correlating
             with a foreground star or a M31 supernova remnant and located 
             in the upper (n$<$1), lower (n$>$1) galaxy hemisphere. Also the 
             distribution of ``intrinsically absorbed'' sources (n$>$2) is 
             shown and the distribution for both hemispheres (0$<$n$<$2). 
             Lower panel: galactocentric distribution of novae (from Sharov \&
             Alksnis 1991, 1992), M31 blue stars with B-V$<$0.3 (from Magnier 
             et al. 1992 and Haiman et al. 1994) and Cepheids (the Baade and 
             Magnier sample, cf. Magnier et al. 1997).}
\end{figure}
 
While novae appear to be bulge-dominated in the observational sample most 
possibly due to the low dust content of the M31 bulge supersoft sources appear
to be to a less degree bulge-dominated and are more likely associated with the
spiral arms. Considering only objects with 0$<$n$<$2 then supersoft sources 
are found within galactocentric radii of 5-25~kpc. They are not found in the 
bulge of M31, but within the range of Cepheids and blue stars and at 
18-22~kpc. Bulge sources at galactocentric radii r$<$6~kpc are only found in 
the ``intrinsically absorbed'' sample. They may be consistent with classical 
or symbiotic novae as these objects show high intrinsic absorption and tend to 
belong to an old population. The M31 supersoft sources may belong to a younger 
population similar to the Cepheids but to an older population than blue stars.
Hatano et al. assume a scale height of the M31 novae of 350~pc. We find that 
the scale height of the M31 supersoft sources is consistent to be smaller 
(100-150~pc, at 5~kpc) which favors a younger stellar population and is in 
agreement with the supported view that slightly evolved main-sequence stars or
subgiants are involved (van den Heuvel et al. 1992). A mean space density can 
be inferred for the population of supersoft sources in M31 (assuming that they
are homogeneously distributed in a disk of radius 20~kpc and have a scale 
height of 150~pc) of $\sim (0.1-5)\times^{-8}\ pc^{-3}$. One could suspect 
from Figure~15 that we see two different sub-populations of supersoft sources,
one in the bulge at galactocentric radii $<$6~kpc, possibly associated with 
(classical and symbiotic) novae, one at radii 12-16~kpc and 18-20~kpc 
respectively tracing spiral arms and possibly associated with subgiants and
CV-type supersoft sources and one at radii 18-22~kpc also associated with a 
spiral arm and possibly associated with subgiants and CV-type supersoft 
sources.

\section{Estimating a SN Ia rate inferred from the population of 
supersoft sources in M31}
Assuming a total number of $\sim$1000-10,000 active supersoft sources in 
M31 as follows from an analysis in paragraph~4.3 and applying the cumulative 
mass distribution from Figure~3 gives a fraction of 26\% in the $t_{\rm 3bol}$
and 32\% in the hydrogen-burning shell approximation for white dwarf masses 
in excess of $0.7\ M_{\odot}$. Assuming that all objects with white dwarf 
masses in excess of $0.7\ M_{\odot}$ explode as type Ia supernovae after a 
typical life time of $10^6$ years, a SN~Ia rate of $\sim(0.3-3)\times 10^{-3}\
{\rm yr^{-1}}$ is inferred (in both approximations). Assuming that all objects
with white dwarf masses in excess of $0.5\ M_{\odot}$ explode as type Ia 
supernovae after a typical life time of $10^6$ years (cf. Yungelson et al. 
1995), a SN~Ia rate of $(0.8-7)\times 10^{-3}\ {\rm yr^{-1}}$ is inferred (for
the two approximations). Supersoft sources could then contribute up to a rate 
of $7\times10^{-3}\ {\rm yr^{-1}}$. Capellaro (1997) assuming our Galaxy to be
a spiral of type Sb or Sc, detected a Type Ia supernova rate of $(2-2.5)\ 
10^{-3}\ {\rm yr^{-1}}$ which for M31, with is about two times larger mass,
means: $(4-5)\ 10^{-3}\ {\rm yr^{-1}}$. It thus seems that the supersoft X-ray 
sources can make a major contribution to the Type Ia SN rate in M31.

In is interesting to note that the historical supernova SN 1885 (S And) might 
be a subluminous SN Ia (Chevalier \& Plait 1988, Fesen et al. 1998). This 
supernova is located in the bulge of M31. 

\section{Conclusions}
   From the 1991 \ros {\sl PSPC} M31 X-ray point source catalog a sample of 
   26 candidate supersoft sources has been derived using one of the 
   selection criteria  $HR1+\sigma HR1 < -0.8$ or $HR1<0.9$, $HR2+\sigma 
   HR2 < -0.1$ and assuming that the observed count rate is in agreement with 
   the expected steady-state luminosity. For these candidates absorbing 
   hydrogen column densities, effective temperatures and white dwarf masses 
   (assuming the sources are on the stability line of atmospheric nuclear 
   burning) are derived. The observed white dwarf mass distribution of
   supersoft sources in M31 appears to be constrained to masses $M\approxgt 
   0.90\ M_{\odot}$. The whole population of supersoft sources in M31 is 
   estimated accordingly to be at least 1000 and at most 10,000 taking a 
   theoretical white dwarf mass distribution and a double exponential scale 
   height distribution for the gas and the source distribution into account 
   and under the assumption that the observationally derived sample is 
   restricted to white dwarf masses above $0.90\ M_{\odot}$. This range of a 
   population has to be compared with the range of a population of 
   $\sim$800-5000 sources as predicted from population synthesis calculations.
   We find the source population scale height to be $\sim$300~pc for a scale 
   height for the gas of 150-600~pc. This is consistent with a young stellar 
   population. Assuming a life time as a steadily nuclear burning white dwarf 
   (a supersoft source) of $\sim10^6\ yr$ and that all supersoft sources with 
   masses in excess of $0.5\ M_{\odot}$ are progenitors of supernovae of type 
   Ia, a SN Ia rate of $\sim (0.8-7)\times 10^{-3}\ {\rm yr^{-1}}$ is inferred
   for M31 based on these progenitors. Supersoft sources then comprise 
   20-100\% of the SNe~Ia progenitors for a total estimated SN Ia rate of 
   $(4-5)\times 10^{-3}\ {\rm yr^{-1}}$.

\acknowledgements
I thank W.Hartmann from SRON for providing the LTE and non-LTE white dwarf 
atmosphere spectra. I thank D. Bhattacharya for discussions. I thank Gene
Magnier for making me the positions of the M31 Cepheids and blue stars
available. I thank E.P.J. van den Heuvel for reading the manuscript. I thank 
an anonymous referee for critical reading of the manuscript and useful 
comments. P.K. is a Human Capital and Mobility fellow. This research was 
supported in part by the Netherlands Organisation for Scientific Research 
(NWO) through Spinoza Grant 08-0 to E.P.J. van den Heuvel.

\end{document}